# Response to the COVID-19 Pandemic: Physics Teaching in India

V. Madhurima[1], Ram Ramaswamy[2], Deepa Chari[3], Vandana Nanal[4, a], and Tanushri Saha-Dasgupta[5]

[1]*Central University of Tamil Nadu*
[2]*Indian Institute of Technology, Delhi*
[3]*Homi Bhabha Centre for Science Education*
[4]*Tata Institute of Fundamental Research*
[5]*S.N. Bose Centre*
[a]nanal@tifr.res.in

**Abstract.** When academic institutions in India closed abruptly in March 2020 due to the COVID-19 pandemic, formal education moved online. This transition had a very uneven impact given the significant digital divide between rural and urban India and the unequal distribution of digital resources in different institutions. Access to resources varied substantially by individual, based on socioeconomic factors as well as gender. Institutional support to the academic community during this critical period was largely inadequate, which has had serious consequences on the teaching of physics and other subjects that require laboratory instruction. Educational institutions also provide safe and enabling learning spaces for women students; reduced access to such facilities undermined the work- and study-related dynamics for women because of the scarcity of resources such as devices, data, and time. This paper reports on efforts made towards understanding such challenges during the COVID-19 pandemic and describes steps that were taken to address them.

## INTRODUCTION

After the declaration of the COVID-19 pandemic in March 2020, education in India moved entirely online. This was a drastic shift for all sectors in education because much (if not all) of school education, as well as science undergraduate and postgraduate teaching, previously relied on physical classrooms and laboratories. Although the peak of the pandemic may have passed, online instruction and pedagogy continue to play a significant role at both the secondary and the tertiary levels of education and will likely also be important in the future. However, the success of the transition to online education has been highly variable owing to the diversity of the country and the uneven access to digital infrastructure. A substantial digital divide exists between rural and urban India, and a large fraction of households in rural sectors are essentially denied access to digital education. Access also depends upon a variety of socioeconomic factors. Not surprisingly, gender is a major factor, with women facing greater risks of early dropouts, career changes, and forced career breaks [1].

The move to online instruction has also reduced the physical access of both teachers and students to the school or university. Campuses can be enabling spaces that provide a sense of belonging and of safety, especially for women students and teachers [2]. Furthermore, within a domestic setting (rather than the classroom) women students and teachers are often burdened with increased familial responsibilities, constrained workspaces, and reduced and uneven access to scarce resources such as devices or data [1, 3].

Beyond these factors, the pandemic period has changed the nature of pedagogy. Institutional support to the academic community (i.e., teachers, teacher educators, students, and researchers) has long fallen short, but it has become seriously inadequate during the pandemic period, with potentially long-term effects on the teaching of physics as well as other subjects for which laboratory instruction is crucial.

It is important to understand the myriad challenges experienced by academic stakeholders due to the rapid switch to the online mode of teaching and learning. This paper describes three national initiatives that explored these issues during the pandemic. An internet group, the Discussion Forum for Online Teaching (DFOT), was set up to address teachers' needs for appropriate resources and practical solutions to aid pedagogy. The national initiatives of Vigyan Pratibha and Vigyan Vidushi reached out to larger teaching-learning communities from school (K–12) to the undergraduate and graduate levels online.

## DISCUSSION FORUM FOR ONLINE TEACHING (DFOT) SURVEY

The absence of a pan-India forum to collaboratively discuss pedagogical concepts prompted the initiation of DFOT [4] in mid-2020. The transition to online teaching in India was quickly recognized as being very uneven, with many teachers finding it difficult to adapt to the new mode of instruction. Most of the problems were local in nature, so one purpose in setting up DFOT was to help teachers share resources, discuss best practices, and offer peer-to-peer training. India has considerable diversity at multiple levels, including economic, social, geographical, and linguistic, and teachers faced multilayered problems. Some problems were pan-Indian (in fact, global), and some were specific to gender or region and economic status. Regular panel discussions were conducted, with experts drawn from all fields of study, different parts of the country, and different types of institutions. The aim was to understand the problems with online teaching and to share solutions with the larger community. A total of 14 panel discussions were held between August 2020 and April 2021; these are all available online [4].

In addition, an online survey was conducted by DFOT, with the questionnaire being shared throughout the country via email and social media. A total of 893 responses were collected between August and December 2020. Details about the respondents were anonymized to protect their privacy, and survey results are available online [8].

The DFOT panel discussions and the survey indicate that owing to inadequate structural support, teachers and students had to use their own personal funds to obtain appropriate resources for online teaching. The DFOT survey indicated a large digital divide between teachers and students as well as within the groups. Access to uninterrupted internet data was a problem for many. While the former was predominantly due to economic status of the family, the latter had a geographical component to it since some parts of India are poorly connected, with several states lacking 3G/4G services. As a result, many students were able to access lectures only on smartphone screens and could not always attend the class without interruption. Survey results are shown in Figure 1.

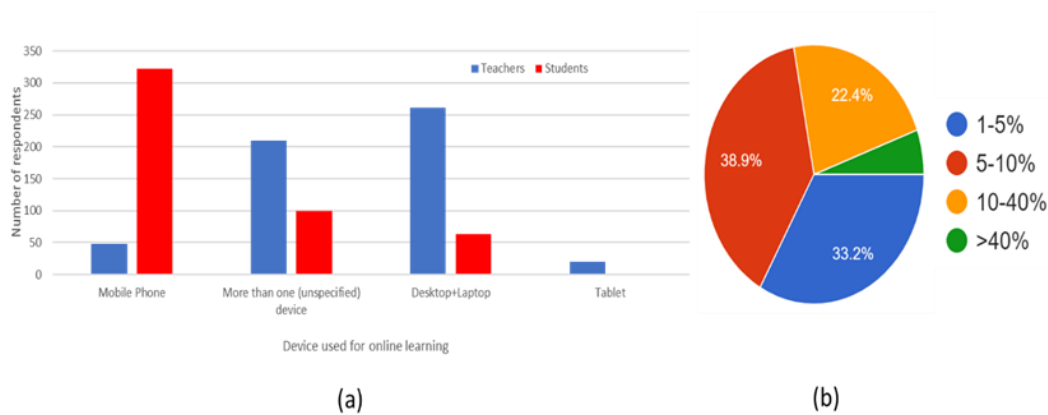

**FIGURE 1.** (a) Type of device(s) used. (b) Proportion of students getting disconnected during a class.

Both students and teachers used a variety of devices to access online classes, namely smartphones, tablets, laptops, and desktops. As shown in Fig. 1a, most teachers used laptops or desktops, while most students used a smartphone. Figure 1b shows the connectivity profile in terms of percentage of classes missed. Although few students missed more than 40% of the classes, a substantial proportion missed at least a few. Reasons for disconnections included loss of signal because of poor connectivity and power-cuts, especially in remote areas of the country.

A parallel effort to collate the responses of a group of teachers and scholars in the tertiary education sector is available online [5]. The reaction of the academic community to the pandemic and to the move online can be assessed from DFOT's survey, the 14 individual DFOT panel discussions [4], and a set of 22 essays about higher education online [5].

The principal findings include the following:
- Although a large number of students log-in to online classes, nearly 40% get disconnected, often several times in one class due to lack of proper internet signals [6–8].
- Most teachers use laptops for teaching, while most students attend classes on (shared) mobile smartphones.
- When mobile phones are shared in a household, girls are at a disadvantage (in terms of access) [1].
- Household work was another area in which women shouldered a larger burden than men [1, 3].
- The anonymity provided by online classes has allowed shy students, usually female, to interact more.
- Conducting physics laboratory courses continues to be a problem. Physics teachers have implemented initiatives such as virtual labs, sending low-cost equipment to the students' homes, virtual labs, and providing students the experimental data to analyze [9].

Programs that encourage young women to take up careers in science, technology, engineering, and mathematics (STEM) disciplines were suspended due to the pandemic [10]. The initiatives described in following sections partially offset these suspensions.

## NATIONAL INITIATIVE OF VIGYAN PRATIBHA FOR SCHOOL SCIENCE

Vigyan Pratibha, which began in 2017, is a student program funded by the Department of Atomic Energy, Government of India and directed by the academic leadership of Homi Bhabha Centre for Science Education (HBCSE), Tata Institute of Fundamental Research (TIFR). The program aims to enhance the scientific and mathematical proficiencies of students in grades 8–10 through building teacher capacity [3, 4].

The crux of the program is a set of expert-developed learning units (LUs) that cover key concepts of science and mathematics. Figure 2 presents an illustration from one of the LUs. The program introduces the LUs and associated pedagogies to high school teachers through residential capacity-building workshops, and the teachers then conduct these activities in their schools. Through participation in these units, students are exposed to the concepts in their textbooks in a richer way. Some common student misconceptions that can hinder understanding are effectively addressed through the LUs. The pandemic halted Vigyan Pratibha's students and teachers camps since March 2020, but the program launched an online teacher professional development program through pedagogy discussion sessions to continue the interaction with science and mathematics teachers across the country.

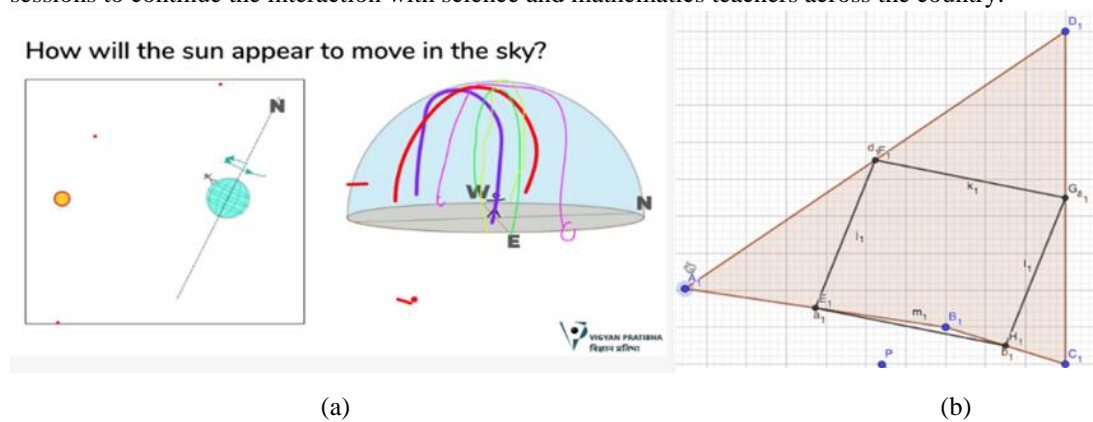

(a) (b)

**FIGURE 2.** (a) A still from "Shadows" learning unit in Vigyan Pratibha. Teachers explore the direction in which Sun will appear to move when seen from the earth's surface standing at a given position using a multiuser board. (b) A still from the "Midpoint Quadrilateral" learning unit session where teachers explored GeoGebra platform to draw and study properties of quadrilaterals.

# VIGYAN VIDUSHI

Vigyan Vidushi (meaning "a learned women scientist" in Sanskrit) is a national level advanced summer school in physics for women students pursuing their first year MSc course [14], jointly organized by TIFR, Mumbai and the HBCSE, TIFR. The objectives of the program are (a) to offer exposure and training to students in advanced physics areas with an emphasis on problem solving and the understanding of core concepts; (b) to provide timely career guidance on opportunities for careers in physics research (globally), and through a group of female physicist mentors; and (c) to expose students to successful women scientists as role models through a series of talks [14]. Additionally, the summer school aims to help students recognize and prepare for potential challenges in physics (e.g., gender bias, imposter syndrome) through a series of workshops and interactions.

The summer school, originally designed as an in-person program, was conducted online, first in July 2020 and again in June 2021 with about 50 students from all across India participating fully in each school in the live online classroom. An example activity is illustrated in Fig. 3a. In addition, over 1,000 students were offered the opportunity to attend lectures on advanced physics topics via livestream. The school included seven core and topical courses in different areas of physics, special lectures by eminent Indian women scientists, virtual lab tours, sessions on Physics Education Research, and career discussion and interactive mentoring sessions (Fig. 3b).

A mixed-gender team of 60 faculty, postdocs, scientific staff, PhD students, and other staff of TIFR main campus and HBCSE participated in different roles as course instructors, tutors, workshop mentors, and support staff. The feedback received from summer school participants after the program indicated that the program was highly successful, and the participants appreciated the interactions with pan-Indian women physicists during the career discussion session. An excerpt of a participant student's feedback is included here.

> *"These interactive sessions were very helpful in the sense that most of the students were not clear about their career options which were cleared by experienced personalities, and the biggest gain was the confidence to speak and share your problem with other people, which was very difficult even for me! In small groups we couldn't hide and have to speak up hence it has boosted our confidence to talk and discuss with others."*

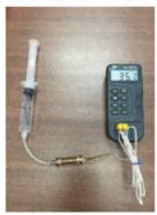
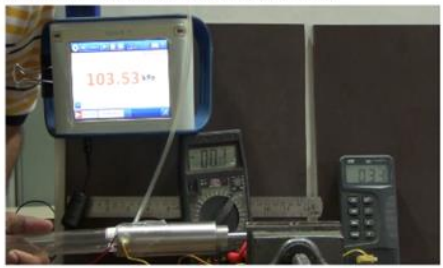

(a)        (b)

**FIGURE 3.** (a) A slide from the experimental techniques session in the Vigyan Vidushi program. (b) A slide from the "Imposter syndrome" closed workshop.

# CONCLUSION

The sudden shift of formal education to the online mode during the pandemic and the closure of educational institutes raised many questions about the preparedness of the academic community—the stakeholders, their coping mechanisms, and issues of access to online education and its reach. Among the initiatives and support groups for education communities that started during the pandemic, DFOT in India provided a forum to discuss and understand problems about access and other pedagogic challenges posed by online teaching. DFOT panel discussions helped teachers cope with the sudden shift by sharing tips and experiences for effective online teaching, evaluation, instruction, and student support. It also provided a platform for peer support through its various social media channels. The pan-Indian survey among teachers conducted by DFOT clearly indicated a need for infrastructural and financial support for online teaching. In the absence of such a support, female students faced the most challenges. It also indicated a need for nationwide initiatives for women students. The other national initiative of Vigyan Pratibha through the online discussion sessions was continued interactions with a limited community of schoolteachers during the pandemic. The Vigyan Vidushi program provided networking and mentoring opportunities for female master's students in physics at an important juncture in their careers. The need for such nationwide initiatives for women students is supported by the results of the DFOT survey. All these online initiatives—although operating for a restricted group of interested student-teacher community—are important since they represent a persistent and continuing effort in STEM teaching and learning, as well as graduate-level mentoring in physics in India.